\documentclass[11pt,twoside]{article}
\usepackage[pdftex]{graphicx}
\usepackage{amsmath}
\usepackage{amssymb}
\usepackage{cite}

\begin{document}
\newcommand{\pst}{\hspace*{1.5em}}


\newcommand{\be}{\begin{equation}}
\newcommand{\ee}{\end{equation}}
\newcommand{\bm}{\boldmath}
\newcommand{\ds}{\displaystyle}
\newcommand{\bea}{\begin{eqnarray}}
\newcommand{\eea}{\end{eqnarray}}
\newcommand{\ba}{\begin{array}}
\newcommand{\ea}{\end{array}}
\newcommand{\arcsinh}{\mathop{\rm arcsinh}\nolimits}
\newcommand{\arctanh}{\mathop{\rm arctanh}\nolimits}
\newcommand{\bc}{\begin{center}}
\newcommand{\ec}{\end{center}}

\thispagestyle{plain}

\label{sh}


\begin{center} {\Large \bf
\begin{tabular}{c}
HIDDEN CORRELATIONS AND ENTANGLEMENT\\[-1mm]
IN SINGLE-QUDIT STATES\footnote{}
\end{tabular}}

\footnotetext{Based on the Talk presented at the QUANTUM 2017 Workshop ad memoriam of Carlo Novero
``From Foundations of Quantum Mechanics to Quantum Information and
Quantum Metrology \& Sensing'' (Turin, Italy, May~7--13, 2017).}

\end{center}

\bigskip

\bc {\bf Margarita A. Man'ko$^1$
and Vladimir I. Man'ko$^{1,\,2}$}

\medskip

{\it $^1$Lebedev Physical Institute, Russian Academy of Sciences\\
Leninskii Prospect 53, Moscow 119991, Russia

\smallskip

$^2$Moscow Institute of Physics and Technology\\
Institutskii per. 9, Dolgoprudny, Moscow Region 141700, Russia}

\smallskip

$^*$Corresponding author e-mail:~~~mmanko\,@\,sci.lebedev.ru

\ec

\begin{abstract}\noindent
We discuss the notion of hidden correlations in classical and
quantum indivisible systems along with such characteristics of the
correlations as the mutual information and conditional information
corresponding to the entropic subadditivity condition and the
entropic strong subadditivity condition. We present an analog of the
Bayes formula for systems without subsystems, study entropic
inequality for von Neumann entropy and Tsallis entropy of the
single-qudit state, and discuss the inequalities for qubit and
qutrit states as an example.
\end{abstract}

\medskip


\section{Introduction}
\pst
The important properties of systems with fluctuations in their
physical characteristics, in both classical and quantum domains,
like positions, momenta, angular momenta, spin, energy, etc., are
connected with correlations in the system degrees of freedom. These
correlations are usually associated with the dependence of the
behavior of the subsystem's degrees of freedom on the behavior of
the other subsystem in composite systems containing two or more
subsystems. In classical domain, this means that one describes the
states of a composite (divisible) system by a joint probability
distribution of several random variables; these aspects are studied
in detail in the probability theory~\cite{Kolmogorov,Holevo-book}.
The notion of entropy is a substantial instrument in the probability
theory~\cite{Shannon,Renyi,Tsallis}. In quantum domain, this means
that one describes the states of such a composite quantum system by
the density operator defined in the Hilbert space~\cite{Dirac-book},
which is the direct product of Hilbert spaces corresponding to the
states of the subsystems, with matrix elements depending on all the
indices corresponding to the basis vectors of these Hilbert spaces.
The mathematical instruments to describe the correlations of the
subsystem degrees of freedom are the Shannon entropy and information
in classical probability theory~\cite{Shannon} and the von Neumann
entropy~\cite{vonNeumannbook} and information in quantum statistics.

The mutual information introduced for classical systems with two
subsystems characterizes the degree of correlations of two
subsystems of the composite system~\cite{Holevo-book}. The quantum
(von Neumann) information of bipartite system, say, the two-qudit
system, provides the characteristics of quantum correlations of the
qudits. For composite systems with three subsystems, the notion of
conditional information is used to describe the degree of
correlations of the subsystems in the composite system. The
mathematical aspects~\cite{Ruskai,Lieb-Ruskai,Rastegin,Ruskai-King}
of these characteristics are presented in terms of
entropy--information inequalities like the subadditivity condition
(bipartite systems) and the strong subadditivity condition
(tripartite systems). The strong quantum correlations in
multipartite systems (qudits) are determined by the entanglement
properties~\cite{Schroedinger,Chuangbook} of the density matrices.

The main idea of our work is to consider noncomposite (nondivisible)
systems, both classical and quantum, and show that there exist
analogs of all correlation properties discussed and all entropies
mentioned above, along with the entropy--information inequalities,
for such systems, as well. For example, the notion of entanglement
and properties of the density matrices of single-qudit states
correspond to specific correlations available in these systems; we
call them the hidden correlations.

It is worth noting that the entanglement of single-qudit states was
discussed for particular examples of the qudits in the
literature~\cite{IJQI14,Shumo,PismaJETPKazan,KassakPhysRev}; also
the notion of generalized entanglement was discussed
in~\cite{Viola1,Viola2,ShumoJPCS}.

In this work, we present a systematic consideration of the
entanglement of the single-qudit states. Also we consider hidden
correlations in qubits and qutrits, using the probability
representation of the density matrices of their states.

Our aim is to discuss hidden correlations available in single-qudit
states~\cite{Entropy,VovaIJQI,MAJRLR13}. The mathematical tools we
are using in this study are based on applying an invertible map of
indices $1,2,\ldots,N$ onto a set of natural indices
$1,2,\ldots,n_1,1,2,\ldots,n_2,\ldots,1,2,\ldots,n_M,$ such that
$N=n_1n_2\cdots n_M$.

The map is realized using the set of ``functions detecting the
hidden correlations''~\cite{ZeilovJRLR2017}. After applying the map,
the possibility arises to interpret an arbitrary probability
distribution of one random variable as a joint probability
distribution of $M$ random variables. In the case of quantum states,
applying this map, one has the possibility to interpret an arbitrary
density matrix of the states in $N$-dimensional Hilbert space as the
density matrix of $M$ qudits. Thus, we extend the notion of
different kinds of quantum correlations known for the qudit-system
($M$ qudits) states to the case of the single-qudit state.

We employ the probability description of qubit and qutrit states,
where the matrix elements of the state density matrices are
explicitly expressed in terms of classical probabilities. Then, in
view of these expressions, we obtain new entropy--information
inequalities for matrix elements of the qubit and qutrit states.

This paper is organized as follows.

In Sec.~2, we describe the invertible map which provides the
possibility to map the density matrix of a single qudit to the
density matrix of the multiqudit system. In Sec.~3, we study the
entanglement of the single-qudit state. In Sec.~4, we present new
entropic inequalities for the qubit and qutrit states. In Sec.~5, we
discuss hidden correlations for spin states in the
tomographic-probability representation. In Sec.~6, we consider an
example of the four-level atom and give the conclusions in Sec.~7.

\section{Partition Map}
\pst
In this section, we describe a map of indices corresponding to the
partition procedure; this map was named the map detecting hidden
correlations in the system\cite{ZeilovJRLR2017}.

Given a set of integers $y=1,2,\ldots,N.$ Consider\,the\,function
$1\geq P(y)\geq0$ satisfying the normalization condition
$\sum_{y=1}^NP(y)=1$. This function can be interpreted as the
probability distribution function of a random variable $y$. If
$N=\prod_{k=1}^MX_k$, one can introduce $M$ functions
$x_1(y),x_2(y),\ldots,x_M(y)$ of the form
\begin{equation}
x_k(y)-1=\frac{y-\left(x_1+\sum_{i=2}^{k}(x_i-1)\prod_{j=1}^{i-1}X_j\right)}
{\prod_{j=1}^{k}X_j}~\mbox{mod}\,{X_k},~~k=1,\ldots,M,~~1\leq y\leq
N.
\label{us}
\end{equation}
Here, the variable $y$ is the function of $x_1,x_2,\ldots,x_M$,
\begin{equation}
y=y(x_1,x_2,\ldots,x_M)=x_1+\sum_{k=2}^{M}(x_k-1)\prod_{j=1}^{k-1}X_j,~~
  1\leq x_k \leq X_ik,~~ k\in[1,M].        \label{su}
  \end{equation}
If $M=2$, the general formulas (\ref{us}) and (\ref{su}) provide
explicit expressions for $y(x_1,x_2)$, $x_1(y)$, and $x_2(y)$
presented as~\cite{ZeilovJRLR2017} (see
also~\cite{Antonella,Facchi})
\begin{eqnarray}
y(x_1,x_2)=x_1+(x_2-1)X_1,\qquad 1\leq x_1 \leq X_1,\quad 1\leq x_2
\leq X_2, \label{map21}\\
 x_1(y)=y \mod{X_1},\qquad 1\leq y \leq N, \label{map22} \\
 x_2(y)-1=\frac{y-x_1(y)}{X_1} \mod{X_2},\qquad 1\leq y \leq
N.\label{map23}
\end{eqnarray}
Functions~(\ref{us}) and (\ref{su}) provide the possibility to
interpret the function $P(y)$ as the joint probability distribution
$\Pi(x_1,x_2,\ldots,x_M)$ of $M$ random variables. We define this
probability distribution by the equality
\begin{equation}
\Pi(x_1,x_2,\ldots,x_M)\equiv P\big(y(x_1,x_2,\ldots,x_M)\big),  \label{1ma}
  \end{equation}
where the argument of the function $P(y)$ is expressed as the
function of $M$ random variables  $x_1,x_2,\ldots,x_M$.

Recall that for any joint probability distribution, one can
introduce marginal probability distributions; for example, if the
integer $s<M$, one has the distribution
\begin{equation}
{\cal P}(x_1,x_2,\ldots,x_s)=\sum_{x_{s+1}=1}^{X_{s+1}}
\sum_{x_{s+2}=1}^{X_{s+2}}
\cdots \sum_{x_{M=1}}^{X_{M}}\Pi (x_1,x_2,\ldots,x_s,x_{s+1},\ldots
x_ M).\label{2ma}
  \end{equation}
In the case $M=2$, one has the joint probability distribution of two
random variables $\Pi(x_1,x_2)$ with marginal probability
distributions ${\cal P}_1(x_1)$ and ${\cal P}_2(x_2)$, which read
\begin{eqnarray}
{\cal P}_1(x_1)=\sum_{x_2=1}^{X_2}P\big(y(x_1,x_2)\big),\qquad
{\cal P}_2(x_2)=\sum_{x_1=1}^{X_1}P\big(y(x_1,x_2)\big).
 \label{4ma}
\end{eqnarray}
Now we consider the conditional probability distribution given by
the Bayes formula (see, e.g.,~\cite{Holevo-book,Bayes})
\begin{equation}
p_1(x_1\mid x_2)=\frac{P\big(y(x_1,x_2)\big)}
{\sum_{x_1=1}^{X_1}P\big(y(x_1,x_2)\big)}.\label{5ma}
  \end{equation}
In view of the map of indices, we can interpret the probability
distribution of one random variable $P(y)$ as a joint probability
distribution of two random variables $x_1$ and $x_2$ and introduce,
along with marginal probability distributions of one random variable
${\cal P}_1(x_1)$ and ${\cal P}_2(x_2)$ given by (\ref{4ma}), two
conditional probability distributions $p_1(x_1\mid x_2)$ given by
(\ref{5ma}) with the other one written as
\begin{equation}
p_2(x_2\mid x_1)=\frac{P\big(y(x_1,x_2)\big)}
{\sum_{x_2=1}^{X_2}P\big(y(x_1,x_2)\big)}.\label{6ma}
  \end{equation}

The generalization to the case $M>2$ is straightforward.

One can introduce the conditional probability distribution of $s<M$
random variables as follows:
\begin{eqnarray}
&&p(x_1,x_2,\ldots,x_s\mid x_{s+1},x_{s+2},\ldots, x_M)\nonumber\\
&&=\frac{P\big(y(x_1,x_2,\ldots,x_s,x_{s+1},\ldots,x_M)\big)} {
\sum_{x_1=1}^{X_1}\sum_{x_2=1}^{X_2}\cdots\sum_{x_s=1}^{X_s}
P\big(y(x_1,x_2,\ldots,x_s,x_{s+1},\ldots,x_M)\big)}\,.
\label{7ma}
\end{eqnarray}

The distributions introduced satisfy all the relationships known for
joint probability distributions, marginal probability distributions,
and conditional probability distributions.

For example, the subadditivity condition for Shannon entropy of
bipartite classical system is expressed as the inequality
$S_1+S_2\geq S(1,2)$ for entropies
\begin{eqnarray}
S_1=-\sum_{x_1=1}^{X_1}{\cal P}_1(x_1)\ln {\cal
P}_1(x_1),\qquad 
S_2=-\sum_{x_2=1}^{X_2}{\cal P}_2(x_2)\ln {\cal
P}_2(x_2),\nonumber\\[-2mm]
\label{9ma}\\[-2mm]
S(1,2)=-\sum_{x_1=1}^{X_1}\sum_{x_2=1}^{X_2}P\big(y(x_1,x_2)\big)\ln
P\big(y(x_1,x_2)\big);\nonumber 
\end{eqnarray}
in an explicit form, the inequality reads
\begin{eqnarray}
&&-\sum_{x_1=1}^{X_1}\left[\sum_{x_2=1}^{X_2}P\big(y(x_1,x_2)\big)\ln
\left(\sum_{x_2=1}^{X_2}P\big(y(x_1,x_2)\big)\right)\right]
-\sum_{x_2=1}^{X_2}\left[\sum_{x_1=1}^{X_1}P\big(y(x_1,x_2)\big)\ln
\left(\sum_{x_1=1}^{X_1}P\big(y(x_1,x_2)\big)\right)\right]\nonumber\\
&&~\qquad\qquad\qquad\qquad\qquad\qquad \geq
-\sum_{x_1=1}^{X_1}\sum_{x_2=1}^{X_2}P\big(y(x_1,x_2)\big)\mid
\ln P\big(y(x_1,x_2)\big).
\label{10ma}
\end{eqnarray}
Here, the function $y(x_1,x_2)$ is given by (\ref{map21}).

\section{Quantum States}
\pst
Now we apply the partition map to the density matrix $\rho_{yy'}$
$(y,y'=1,2,\ldots,N)$ of the single-qudit state either to the $N$
level atom or to the spin-$j$ state, where $N=2j+1$. The density
matrix $\rho_{yy'}$ is a nonnegative Hermitian matrix
$\rho=\rho^\dagger$, $\rho\geq 0$, with unit trace Tr$\,\rho=1$. The
nonnegativity of the matrix means that the eigenvalues of the matrix
$\lambda_1,\lambda_2,\ldots,\lambda_N$ are nonnegative numbers,
i.e., $\lambda_k\geq 0$ $(k=1,2,\ldots,N)$. The density matrix
$\rho_{yy'}$ can be considered as the $N$$\times$$N$ matrix $R$ with
matrix elements defined by the equality
\begin{equation}\label{11ma}
R_{x_1,x_2,\ldots,x_M,x'_1,x'_2,\ldots,x'_M}=
\rho_{y(x_1,x_2,\ldots,x_M),y'(x'_1,x'_2,\ldots,x'_M)}.
\end{equation}
Numerically both matrices are identical, but the interpretation of
the matrix $R$ is different from the interpretation of the matrix
$\rho$.

The matrix $\rho$ is interpreted as the density matrix of the
single-qudit state. The density matrix $R$ is interpreted as the
density matrix of the multiqudit state. This fact means that the
same numerical matrices $R$ and $\rho$ can be interpreted either as
the density matrix of the state of a system without subsystems or as
the density matrix of the state of a system with $M$ subsystems.

For a composite system with the density matrix $R$, one has the
density matrices of the states of the subsystems; these density
matrices are obtained by applying the partial tracing procedure. One
can apply this tool to associate with the density matrix $\rho$ of
the single-qudit state the density matrix $\rho(1)$ with matrix
elements defined by an analogous partial tracing tool, in view of
the formula
\begin{eqnarray}
&&\rho_{x_1,x_2,\ldots,x_s,x'_1,x'_2,\ldots,x'_s}(1) \nonumber\\
&& =\sum_{x_{s+1=1}}^{X_{s+1}}
\sum_{x_{s+2=1}}^{X_{s+2}}\cdots\sum_{x_M=1}^{X_M}
\rho_{y(x_1,x_2,\ldots,x_s,x_{s+1},x_{s+2},\ldots,x_M),
y'(x'_1,x'_2,\ldots,x'_s,x_{s+1},x_{s+2},\ldots,x_M)}. \label{12ma}
\end{eqnarray}
The density matrix $\rho(1)$ is interpreted as the density matrix of
an artificial subsystem state of $s$ qudits.

Analogously, one can introduce the density matrix $\rho(2)$ of an
artificial subsystem state of $M-s$ qudits using the definition
formulated due to the other partial tracing procedure,
\begin{eqnarray}
&&\rho_{x_{s+1},x_{s+2},\ldots,x_M,x'_{s+1},x'_{s+2},\ldots,x'_M}(2)\nonumber\\
&&=\sum_{x_1=1}^{X_1}
\sum_{x_2=1}^{X_2}\cdots\sum_{x_s=1}^{X_s}
\rho_{y(x_1,x_2,\ldots,x_s,x_{s+1},x_{s+2},\ldots,x_M),
y'(x_1,x_2,\ldots,x_s,x'_{s+1},x'_{s+2},\ldots,x'_M)}.\label{13ma}
\end{eqnarray}

The von Neumann entropy of the state of a noncomposite system with
the density matrix $\rho$, i.e.,
$~S=-\mbox{Tr}\,\rho\ln\rho\equiv-\mbox{Tr}\,R\ln R$, is a
nonnegative number $S\geq 0$.

Using the introduced partition map of indices and the introduced
subsystem density matrices of artificial subsystem states, one can
define the notion of mutual quantum information $I_q$ using the
standard relation for composite systems. In the explicit form, the
nonnegative mutual quantum information reads
\begin{equation}\label{14ma}
I_q=\mbox{Tr}\left(\rho\ln\rho\right)-\mbox{Tr}\left(\rho(1)\ln\rho(1)\right)
-\mbox{Tr}\left(\rho(2)\ln\rho(2)\right).
\end{equation}
If in artificial subsystems 1 and 2 associated with the single-qudit
state with the density matrix $\rho$, there are no correlations,
i.e., $\rho=\rho(1)\otimes\rho(2)$, then the mutual information
$I_q=0$. The difference of information from zero corresponds to the
presence of correlations of artificial subsystems~1 and 2 in the
single-qudit system. We call these correlations ``hidden
correlations.''

The notion of entanglement for the single qudit state is defined
analogously to the case of the entanglement definition in the
multiqudit systems. If the density matrix $\rho$ of the single-qudit
state can be presented in the form of convex sum,
\begin{eqnarray}
\rho_{y(x_1,x_2,\ldots,x_M),y'(x'_1,x'_2,\ldots,x'_{M})}
=\sum_kp_k\,r^{(k)}_{x_1,x_2,\ldots,x_s,x'_1,x'_2,\ldots,x'_s}(1)
\times r^{(k)}_{x_{s+1},x_{s+2},\ldots,x_M,x'_{s+1},x'_{s+2},\ldots,x'_M}(2),
\label{14mam}
\end{eqnarray}
where $r^{(k)}(1)$ and $r^{(k)}(2)$ are density matrices of the
artificial subsystem states, the single-qudit state is called the
separable state. In~(\ref{14mam}), $1\geq p_k\geq 0$ and
$\sum_kp_k=1$. If the density matrix $\rho_{yy'}$ cannot be
presented in the form~(\ref{14mam}), the single-qudit state is
called the entangled state. The entangled state of a single qudit
interpreted as a collection of multiqudit subsystems is the state
with strong quantum correlations of artificial subsystems. For
example, if $M=2$, one has the separability condition for the
density matrix $\rho_{yy'}$ of the single-qudit state as follows:
\begin{equation}\label{15ma}
\rho_{y(x_1,x_2),y'(x'_1,x'_2)}=\sum_kp_k\,r^{(k)}_{x_1,x'_1}(1)\times
r^{(k)}_{x_2,x'_2}(2).
\end{equation}
For the pure state of a single qudit, the entanglement can be
characterized by the entropy of an artificial subsystem state
analogously to the entanglement of the pure state of the composite
system. One has the parameter called the linear entropy,
\begin{equation}\label{16ma}
E=1-\mbox{Tr}\,\big(\rho(2)\big)^2;
\end{equation}
it is a number equal to zero if there is no hidden correlations of
the artificial subsystems.

\section{New Entropic Inequalities for the Density Matrix
Elements of Qubit and Qutrit States}
\pst
In this section, we present some examples of new entropic
inequalities for density matrices of the qubit and qutrit states
using the tomographic probabilities determining the density
matrices. We express the density matrix element $\rho_{1/2,-1/2}$ of
the qubit state in terms of probabilities $p_1$ and $p_2$ of the
spin projections $m=1/2$ on the $x$ and $y$~axes as
follows~\cite{VitaleJPA2017,VovaJRLR-2-2017}:
\begin{equation}
\rho_{1/2,-1/2}=\langle 1/2\mid\hat\rho\mid
-1/2\rangle=p_1-ip_2-(1-i)/2,
\label{1}
\end{equation}
and the matrix element $\rho_{1/2,1/2}$ of this matrix is expressed
in terms of the probability of the spin projection $(m=1/2)$ equal
to $p_3$ on the $z$~axis as $\rho_{1/2,1/2}=p_3.$

Since three classical probability distributions $(p_1, 1-p_1)$,
$(p_2, 1-p_2)$, and $(p_3, 1-p_3)$ satisfy the condition of
nonnegativity of the relative Shannon~\cite{Shannon} and
Tsallis~\cite{Tsallis} entropies, the following entropic
inequalities for the matrix elements of the qubit density matrices
hold:
\begin{eqnarray}
&&~~\left[({1}/{2})+\mbox{Re}\,\rho_{1/2,-1/2}\right]\ln\left[
\left[({1}/{2})+\mbox{Re}\,\rho_{1/2,-1/2}\right]\left(\rho_{1/2,1/2}\right)^{-1}\right]\nonumber\\
&&+\left[({1}/{2})-\mbox{Re}\,\rho_{1/2,-1/2}\right]\ln\left[
\left[({1}/{2})-\mbox{Re}\,\rho_{1/2,-1/2}\right]\left(\rho_{-1/2,-1/2}\right)^{-1}\right]\geq
0,\label{3}\\
&&~~\left[({1}/{2})+\mbox{Re}\,\rho_{1/2,-1/2}\right]\ln\left[
\left[({1}/{2})+\mbox{Re}\,\rho_{1/2,-1/2}\right]\left(( {1}/{2})+\mbox{Im}\,\rho_{1/2,-1/2}\right)^{-1}\right]\nonumber\\
&&+\left[({1}/{2})-\mbox{Re}\,\rho_{1/2,-1/2}\right]\ln\left[
\left[({1}/{2})-\mbox{Re}\,\rho_{1/2,-1/2}\right]\left(({1}/{2})-\mbox{Im}\,\rho_{1/2,-1/2}\right)^{-1}\right]\geq
0.\label{4}
\end{eqnarray}
These two inequalities are compatible with the nonnegativity
condition of the qubit density matrix.

The qutrit-state density matrix obtained in terms of the nine
probabilities, satisfying the constrain $~1\geq p_1^{(1)}$,
$p_2^{(1)}$, $p_3^{(1)}$, $p_1^{(2)}$, $p_2^{(2)}$, $p_3^{(2)}$,
$p_1^{(3)}$, $p_2^{(3)}$, $p_3^{(3)}\geq 0$ and corresponding to
three artificial qubits describing the qutrit state, has the matrix
elements~\cite{VovaJRLR-2-2017}
\begin{equation}\label{5}
\rho_{11}=p_3^{(2)}+p_3^{(1)}-1,\quad
\rho_{22}=1-p_3^{(2)},\quad \rho_{21}=p_1^{(2)}+ip_2^{(2)}-(i+1)/2.
\end{equation}
Since the matrix elements are expressed in terms of classical
probability distributions, the nonnegativity condition for relative
entropy yields new inequalities for the density matrix elements of
the qutrit state; one of the inequalities reads
\begin{eqnarray}
(\rho_{11}+\rho_{22})\ln\left\{(\rho_{11}+\rho_{22})\big[(1/2)
+\mbox{Re}\,\rho_{13}\big]^{-1}\right\}
+\rho_{33}\ln\left\{(\rho_{33}+\rho_{22})\big[(1/2)
-\mbox{Re}\,\rho_{13}\big]^{-1}\right\}\geq 0.\label{3a}
\end{eqnarray}
In view of the nonnegativity condition of the Tsallis relative
entropy, we obtain a new inequality for the density matrix elements
of the qutrit state; for $q>1$, it is
\begin{eqnarray}
-(1-q)^{-1}\left\{(\rho_{11}+\rho_{22})^q\big[(1/2)
+\mbox{Re}\,\rho_{13}\big]^{1-q}\right. +\rho_{33}^q\big[(1/2)\left.
-\mbox{Re}\,\rho_{13}\big]^{1-q}-1\right\}\geq 0.
\label{4a}
\end{eqnarray}
The entropic inequalities obtained reflect the presence of hidden
correlations of artificial qubits associated with the qutrit-state
density matrix. (Analogous inequalities can be found for qudit
states.) The inequalities are compatible with the inequalities like
the nonnegativity condition of the qutrit-state density matrix and
the nonnegativity of the von Neumann entropy
$~S=-\mbox{Tr}\,(\rho\ln\rho)\geq 0$. But the inequalities obtained
are new; they can be checked in the experiments, where the
qutrit-state tomography provides the reconstruction of the matrix
elements of the density matrix, e.g., in superconducting circuits
based on Josephson
junctions~\cite{OlgaJSLR1989,Ast,Ust,Walral,Devoret,Kiktenko,Glushkov,Fujii}.

\section{Hidden Correlations for Tomographic-Probability
Distributions of Spin-${\boldsymbol j}$ States}
\pst
Applying the partition map, we obtain new entropic inequalities for
tomographic-probability distribution $w(m\mid\vec n)$\cite{Beauty}
describing the single spin-$j$ states; here, $m$ is the spin-$j$
projection on the direction, given by the unit vector $\vec n$, to
be equal $m$. The tomogram is defined~\cite{OlgaJETP,DodonPLA} (see
also~\cite{Weigert,Amiet}) as the diagonal matrix element of the
density matrix $\rho_{mm}(\vec n)$, with $\vec
n=(\cos\varphi\sin\theta,\sin\varphi\sin\theta,\cos\theta)$, and the
matrix element expressed in terms of the spin-state density operator
$\hat\rho$ and the unitary operator $\hat u$ of the irreducible
representation of the group $SU(2)$ by the formula
\begin{equation}\label{A1}
w(m\mid\vec n)=\langle m\mid\hat u\hat\rho\hat u^\dagger\mid
m\rangle,
\end{equation}
where $\mid m\rangle$ is the eigenvector of the spin projection
operator $\hat J_z$ on the $z$~direction, i.e., $\hat J_z\mid
m\rangle=m\mid m\rangle$, and the operator $\hat u$ matrix elements
$\langle m\mid\hat u\mid m'\rangle=u_{mm'}$ are the functions of the
Euler angles~\cite{LanLifQuantMech}.

For spin $j=1/2$, the 2$\times$2~matrix $u_{mm'}$ reads
\begin{equation}\label{A2}
u_{mm'}(\varphi,\theta,\psi)=\left(\begin{array}{cc}
\cos(\theta/2)\,e^{i(\psi+\varphi)/2} & ~~\sin(\theta/2)\,e^{i(\psi-\varphi)/2}  \\
-\sin(\theta/2)\,e^{-i(\psi-\varphi)/2} &
~~\cos(\theta/2)\,e^{-i(\psi+\varphi)/2}\end{array}\right).
\end{equation}
In view of the structure of (\ref{A1}), the tomogram $w(m\mid\vec
n)$ depends only on two angles $\varphi$ and $\theta$ determining
the unit vector $\vec n$.

For given vector $\vec n$, the tomogram is the normalized
conditional probability distribution satisfying the equality
\begin{equation}\label{A3}
\sum_{m=-j}^jw(m\mid\vec n)=1,\qquad m=-j,-j+1,\ldots, j-1,j.
\end{equation}
The tomogram determines the density operator $\hat\rho$ of a single
qudit. It is worth noting that the symplectic tomography for
continuous variables was also
introduced~\cite{ManciniPLA,IbortPS150}.

To obtain a new relationship for tomogram, we introduce the new
notation $w(m\mid\vec n)\equiv P(y\mid\vec n)$, where $y(-j)=1$,
$y(-j+1)=2$, $y(j-1)=N-1$, and $y(j)=N=(2j+1)$. In view of this
change of the variables, we obtain $m=-j\to 1,m=-j+1\to
2,\ldots,m=j-1\to N-1,m=j\to N$.

We consider the spin tomogram as the probability distribution of one
random variable $y$ discussed in the previous sections. In this way,
we can obtain new entropic inequalities describing hidden
correlations for the single qudit (spin-$j$) system. If spin $j$ of
the system is such that $2j+1=X_1X_2$, i.e., in previous
formulas~(\ref{su}) and (\ref{4ma}), we have $M=2$,
$x_1=1,2,\ldots,X_1$, and $x_2=1,2,\ldots,X_2$, one can apply the
partition map and introduce two artificial subsystems corresponding
to the probability distributions
\begin{eqnarray}
{\cal P}_1(x_1\mid\vec n)=\sum_{x_2=1}^{X_2}w\big(m\to
y(x_1,x_2)\mid\vec n\big),\qquad {\cal P}_2(x_2\mid\vec
n)=\sum_{x_1=1}^{X_1}w\big(m\to y(x_1,x_2)\mid\vec n\big).\label{A5}
\end{eqnarray}
Using the definition of Tsallis entropy~\cite{Tsallis}
\begin{eqnarray}
&&S_q^{(1)}(\vec n_1)=\frac{1}{1-q}\left[\sum_{x_1=1}^{X_1}{\cal
P}_1^q(x_1\mid\vec n_1)-1\right],\qquad S_q^{(2)}(\vec
n_2)=\frac{1}{1-q}\left[\sum_{x_2=1}^{X_2}{\cal P}_2^q(x_2\mid\vec
n_2)-1\right],\nonumber\\[-2mm]
&&\label{A6}\\[-2mm]
&&S_q(\vec n)=\frac{1}{1-q}\left[\sum_{x_1=1}^{X_1}
\sum_{x_2=1}^{X_2}w^q\big(m\to y(x_1,x_2)\mid\vec n\big)-1\right]
=\frac{1}{1-q}\left[\sum_{m=-j}^jw(m\mid\vec n)^q-1\right]
\nonumber
\end{eqnarray}
and the known nonnegativity of information along with the conditions
for relative Tsallis entropy, we arrive at new inequalities for spin
tomograms of quantum states; they read
\begin{eqnarray}
S_q^{(1)}(\vec n)+S_q^{(2)}(\vec n)\geq S_q(\vec n),\label{A7}\\
\frac{1}{q-1}\left[\sum_{x_k=1}^{X_k}{\cal
P}_1^q(x_k\mid\vec n_1){\cal P}_2^{1-q}(x_k\mid\vec
n_2)-1\right]\geq 0,\quad k=1,2,\quad X_1=X_2.\label{A8}
\end{eqnarray}
For $q\to 1$, the Tsallis entropic relations yield the relations for
the Shannon entropy. For example, for $q=1$, Eq.~(\ref{A7}) provides
the nonnegativity condition for the mutual tomographic information
\begin{eqnarray}\label{A9}
I(\vec n)&=&-\sum_{x_1=1}^{X_1}\left[\sum_{x_2=1}^{X_2}w\big(m\to
y(x_1,x_2)\mid\vec
n\big)\right]\ln\left[\sum_{x_2=1}^{X_2}w\big(m\to
y(x_1,x_2)\mid\vec n\big)\right]\nonumber\\
&&-\sum_{x_2=1}^{X_2}\left[\sum_{x_1=1}^{X_1}w\big(m\to
y(x_1,x_2)\mid\vec
n\big)\right]\ln\left[\sum_{x_1=1}^{X_1}w\big(m\to
y(x_1,x_2)\mid\vec n\big)\right]\nonumber\\
&&+\sum_{x_1=1}^{X_1}\sum_{x_2=1}^{X_2}w\big(m\to y(x_1,x_2)\mid\vec
n\big)\ln w\big(m\to y(x_1,x_2)\mid\vec n\big)\geq 0.
\end{eqnarray}
If the hidden correlations are not present, information $I(\vec
n)=0$, and if information $I(\vec n)$ is large, the hidden
correlations in the state with tomogram $w(m\mid\vec n)$ are strong.

\section{Examples of the Spin-3/2 System and the Four-Level Atom}
\pst
In this section, we study in detail the hidden correlations in the
four-level atomic system; this means that we consider also the
particle with spin $j=3/2$. The states of this particle (single
qudit) are associated with vectors $\mid -3/2\rangle$, $\mid
-1/2\rangle$, $\mid 1/2\rangle$, and $\mid 3/2\rangle$. The states
of the four-level atom are associated with vectors $\mid 1\rangle$,
$\mid 2\rangle$, $\mid 3\rangle$, and $\mid 4\rangle$, which are the
energy eigenvectors with energies $E_1$,  $E_2$,  $E_3$, and $E_4$.

The spin-3/2 states are the eigenstates $\mid m\rangle$ of the
spin-projection operator $\hat J_z$, i.e.,  $\hat J_z\mid
m\rangle=m\mid m\rangle$. The spin projection on the $z$ axis can
take values $m=-3/2,-1/2,1/2,3/2$. The Hilbert space of the
discussed system states is four-dimensional. The density matrix
$\rho_{mm'}$ of any state in this space for the spin-3/2 particle
reads
\begin{equation}\label{T1}
\rho=\left(\begin{array}{cccc}
\rho_{-3/2~-3/2}&~\rho_{-3/2~-1/2}&~\rho_{-3/2~1/2}&~\rho_{-3/2~3/2}\\
\rho_{-1/2~-3/2}&~\rho_{-1/2~-1/2}&~\rho_{-1/2~1/2}&~\rho_{-1/2~3/2}\\
\rho_{1/2~-3/2}&~\rho_{1/2~-1/2}&~\rho_{1/2~1/2}&~\rho_{1/2~3/2}\\
\rho_{3/2~-3/2}&~\rho_{3/2~-1/2}&~\rho_{3/2~1/2}&~\rho_{3/2~3/2}\end{array}\right).
\end{equation}
For the four-level atom, the same numerical density matrix
$\rho_{nn'}$ is
\begin{equation}\label{T2}
\rho=\left(\begin{array}{cccc}
\rho_{11}&\rho_{12}&\rho_{13}&\rho_{14}\\
\rho_{21}&\rho_{22}&\rho_{23}&\rho_{24}\\
\rho_{31}&\rho_{32}&\rho_{33}&\rho_{34}\\
\rho_{41}&\rho_{42}&\rho_{43}&\rho_{44}
\end{array}\right).
\end{equation}
Thus, in our formalism of the partition map, we have $N=4$, $M=2$,
and $X_1=X_2=2$, along with the numbers $x_1=1,2$ and $x_2=1,2$.

These different physical systems (four-level atom and spin-3/2
particle) model the qudit with the same numerical density matrix.
The invertible map of the spin-density matrix onto the
four-level-atom density matrix uses the change of indices $-3/2\to
1$, $-1/2\to 2$, $1/2\to 3$, and  $3/2\to 4$.
Formulas~(\ref{map21})--(\ref{map23}) provide the numerical values
for the functions $x_1(y)$, $x_2(y)$, and $y(x_1,x_2)$; they are
\begin{eqnarray}
y(1,1)=1,\quad y(2,1)=2,\quad y(1,2)=3,\quad y(2,2)=4,\label{T3}\\
x_1(1)=1,\quad x_1(2)=2,\quad x_1(3)=1,\quad x_1(4)=2,\label{T4}\\
x_2(1)=1,\quad x_2(2)=1,\quad x_2(3)=2,\quad x_2(4)=2.\label{T5}
\end{eqnarray}
Thus, for the four-level-atom density matrix~(\ref{T2}) and the same
numerical elements, we have the expression
$\rho\big(y(x_1,x_2)y'(x'_1,x'_2)\big)\equiv R_{x_1x_2,x'_1x'_2}$.
The form of this matrix coincides with the density matrix of the
two-qubit system: The states of the first and second artificial
qubits have the density matrices in terms of matrix elements
$\rho_{nn'}$, namely,
\begin{equation}\label{T6}
\rho(1)=\left(\begin{array}{cc}
\rho_{11}+\rho_{22}&\rho_{13}+\rho_{24}\\
\rho_{31}+\rho_{42}&\rho_{33}+\rho_{44}
\end{array}\right),\qquad
\rho(2)=\left(\begin{array}{cc}
\rho_{11}+\rho_{33}&\rho_{12}+\rho_{34}\\
\rho_{21}+\rho_{43}&\rho_{22}+\rho_{44}
\end{array}\right).
\end{equation}
The two artificial qubit states determine quantum information
\begin{equation}\label{T7}
I=-\mbox{Tr}\,\rho(1)\ln\rho(1)-\mbox{Tr}\,\rho(2)\ln\rho(2)
+\mbox{Tr}\,\rho\ln\rho;
\end{equation}
it is equal to zero if there is no hidden correlations of artificial
qubits in the four-level atom. The entangled states of the
spin-3/2\, particle\, can be\, presented as\,
$\mid\psi\rangle=2^{-1/2}(\mid 3/2\rangle+\mid -3/2\rangle)$. One
can check that for the two-qubit states, this state is the Bell
state. The Bell inequality written for two artificial qubits in this
spin-3/2 state is violated.

\section{Conclusions}
\pst
To conclude, we formulate the main results of our work.

We showed that there exist correlations in noncomposite
(nondivisible, both classical and quantum) systems which are known
for multipartite systems. We obtained entropy--information
inequalities which are new relations for the systems without
subsystems. We formulated the notion of entanglement for
single-qudit states. In view of the partition map of indices
labeling the matrix elements of the density matrices, we described
systematically the results obtained. The new entropic inequalities
obtained for the qubit~(\ref{3}),~(\ref{4}) and
qutrit~(\ref{3a}),~(\ref{4a}) states can be checked in experiments
with superconducting circuits based on Josephson-junction devices.

We showed that the known entropic inequalities which are applied to
composite systems, both classical and quantum, can also be applied
to the systems without subsystems. In view of the interpretation of
the density matrix of the noncomposite system as the density matrix
of an artificial bipartite system, we obtained a new entropic
inequality~(\ref{A9}) for the qudit spin tomogram.

In fact, the approach presented provides the possibility to extend
all entropic and information relations known for for classical and
quantum composite systems to the case of the systems without
subsystems; these relations reflect the presence of correlations,
either classical or quantum, of the system's degrees of freedom. The
quantum correlations of the single-qudit
states~\cite{NC-2013,PS2014,physconf2013,JRLR2014,JRLR2015,QuantumFest,NC-2016,JRLR2016,JRLR2017,MarkJRLR2017}
can be used for quantum technologies analogously to the employment
of entanglement as a quantum resource.

\section*{Acknowledgments}
\pst
We are grateful to the Organizers of the QUANTUM 2017 Workshop ad
memoriam of Carlo Novero ``From Foundations of Quantum Mechanics to
Quantum Information and Quantum Metrology \& Sensing'' (Turin,
Italy, May~7--13, 2017) and especially to Prof. Marco Genovese for
invitation and kind hospitality.

\end{document}